\documentclass{svmult}
\usepackage{makeidx}  
\usepackage{graphicx}  
\usepackage{subfigure}   
\usepackage{epsfig}          
\usepackage{mathptmx,helvet,courier}  
\usepackage{multicol}        
\usepackage[bottom]{footmisc} 
\usepackage{cite}
\usepackage{amssymb} \usepackage{amsmath} \usepackage{amsfonts}

\def\rcgindex#1{\index{#1}}
\def\myidxeffect#1{{\bf\large #1}}

\begin{document}

 \title*{Rate theory of acceleration of  defect annealing driven by discrete breathers} 
  \titlerunning{Rate theory of acceleration of  defect annealing driven by discrete breathers}
 \author{Vladimir~I.~Dubinko \and Juan~F.~R.~Archilla \and Sergey~V.~Dmitriev \and  Vladimir~Hizhnyakov
 } 
\institute{
  V.I.~Dubinko
  \at NSC Kharkov Institute of Physics and Technology, Akademicheskya Str.~1, Kharkov 61108,
  Ukraine, \email{vdubinko@mail.ru}
  \and J.F.R.~Archilla
  \at Group of Nonlinear Physics, Universidad de Sevilla, ETSI Inform\'atica,
  Avda. Reina Mercedes s/n 41011, Seville, Spain,
  \email{archilla@us.es}
  \and S.V.~Dmitriev
  \at Institute for Metals Superplasticity Problems, RAS, Khalturin Str. 39, Ufa 450001, Russia \\
  National Research Tomsk State University, Lenin Prospekt 36, Tomsk 634050,
  Russia, \email{dmitriev.sergey.v@gmail.com}
  \and V.~Hizhnyakov
  \at Institute of Physics, University of Tartu, Riia 142, EE-51014 Tartu,
  Estonia, \email{hizh@fi.tartu.ee}
  } 
\authorrunning{V.I.~Dubinko, J.F.R. Archilla, S.V.~Dmitriev  and  V.~Hizhnyakov}
\tocauthor{V.I.~Dubinko, J.F.R. Archilla, S.V.~Dmitriev  and
V.~Hizhnyakov}
\providecommand{\degC}{$^\circ$C}
\maketitle \vspace{-2cm} \setcounter{minitocdepth}{1} \dominitoc


\abstract{Novel mechanisms of defect annealing in solids are
discussed, which are based on the large amplitude anharmonic
lattice vibrations, a.k.a. intrinsic localized modes or discrete
breathers (DBs). A model for amplification of defect annealing
rate in Ge by low energy plasma-generated DBs is proposed, in
which, based on recent atomistic modelling, it is assumed that DBs
can excite atoms around defects rather strongly, giving them
energy $\gg k_BT$ for $\sim$100 oscillation periods. This is shown
to result in the amplification of the annealing rates proportional
to the DB flux, i.e. to the flux of ions (or energetic atoms)
impinging at the Ge surface from inductively coupled plasma (ICP).
 }

\section{Introduction}
\label{sec:dubinko:introduction}
 A defect lying in the band gap with
energy $> 0.1$ eV from either band edge is termed deep. As known
from the studies of properties of defects in Ge
\cite{rate-auret2010,rate-auret2007,rate-markevich2004,rate-archillacoelho2015,%
 rate-archillacoelhoquodons2015},
  Ar ions arriving at a semiconductor
surface with very low energy (2 - 8 eV) are annihilating defects
deep inside the semiconductor. Several different defects were
removed or modified in Sb-doped germanium, of which the $E$-center
has the highest concentration, as described in details in
Ref.~\cite{rate-archillacoelho2015,rate-archillacoelhoquodons2015}. 
 Novel mechanisms of defect annealing in solids are discussed in
this work, which are based on the large amplitude anharmonic
lattice vibrations, a.k.a. intrinsic localized modes (ILMs) or
discrete breathers (DBs). The article is organized as follows. In
Sect.~\ref{sec:dubinko:DBreview}, a short review on DB properties
in metals and semiconductors is presented based on the results of
molecular dynamics (MD) simulations using realistic many-body
interatomic potentials. In Sect.~\ref{sec:dubinko:DBexcitation}, a
rate theory of DB excitation under thermal heating and under
non-equilibrium gas loading conditions is developed. In
Sect.~\ref{sec:dubinko:DBamplification}, a model for amplification
of defect annealing rate in Ge by plasma-generated DBs is proposed
and compared with experimental data. The results are summarized in
Sect.~\ref{sec:dubinko:summary}.

\section{Discrete breathers in metals and semiconductors}
\label{sec:dubinko:DBreview} 

DBs are spatially localized large-amplitude vibrational modes in
lattices that exhibit strong anharmonicity
  \cite{rate-sievers1988,rate-hizhnyakov2002,rate-piazza2003,rate-flach2008}. 
 They have been identified as exact solutions to a number of model
nonlinear systems possessing translational symmetry
\cite{rate-flach2008} and successfully observed experimentally in
various physical systems \cite{rate-flach2008,rate-manley2010}.
Presently the interest of researchers has shifted to the study of
the role of DBs in solid state physics and their impact on the
physical properties of materials
 \cite{rate-manley2010,rate-dubinkorussell2011,
 rate-dubinkoarchilla2011,rate-dubinkodubinko2013,
 rate-dubinkospringer2014,rate-terentyevdubinko2014}.
        Until recently the evidence for the DB
existence provided by direct atomistic simulations, e.g. MD, was restricted mainly to one and two-dimensional
networks of coupled nonlinear oscillators employing oversimplified
pairwise inter-particle potentials
   \cite{rate-hizhnyakov2002,rate-piazza2003,rate-flach2008}. 
Studies of the DBs in three-dimensional systems by means of MD simulations using
realistic interatomic potentials include ionic crystals with NaCl
structure
  \cite{rate-kiselev1997,rate-khadeeva2010}, 
   graphene \cite{rate-khadeevakivshar2011,rate-baimovaZhou2012,rate-korznikova2013},     
   graphane \cite{rate-liu2013},     
   semiconductors \cite{rate-voulgarakis2004}, 
pure metals
\cite{rate-haas2011,rate-hizhnyakov2014,rate-terentyevdubinko2014,rate-murzaev2015},
and
ordered alloys \cite{rate-medvedev2013}. 
For the first time the density functional theory (DFT) was applied
to the study of DBs, using graphane as an example
\cite{rate-chechin2014}.

DBs have very long lifetime because their frequencies lie outside
the phonon band. Monatomic crystals like pure metals and
semiconductors such as Si and Ge do not possess gaps in the phonon
spectrum, while crystals with complex structure often have such
gaps, for example, diatomic alkali halide crystals and ordered
alloys with a large difference in the atomic mass of the
components. For the crystals possessing a gap in the phonon
spectrum the so-called gap DBs with frequencies within the gap can
be excited. This case will not be discussed here and in the
following we focus on the DBs having frequencies above the phonon
band. \rcgindex{\myidxeffect{G}!Gap breathers}
\rcgindex{\myidxeffect{B}!Breathers in the phonon gap}

\subsection{Metals} \label{subsec:dubinko:metals} 

In the work by Kiselev {\em et al.} \cite{rate-kiselev1993} it has
been demonstrated that 1D chain of particles interacting with the
nearest neighbors via classical pairwise potentials such as Toda,
Lennadrd-Jones or Morse cannot support DBs with frequencies above
the phonon band. They were able to excite only gap DBs with
frequencies lying within the gap of the phonon spectrum by
considering diatomic chains. In line with the results of this
work, it was accepted for a long time that the softening of atomic
bonds with increasing vibrational amplitude is a general property
of crystals, which means that the oscillation frequency decreases
with increasing amplitude. Therefore DBs with frequencies above
the top phonon frequency were unexpected.

However, in 2011, Haas et al.~\cite{rate-haas2011} 
  have demonstrated by MD simulations using realistic many-body interatomic
potentials that DBs with frequencies above the phonon spectrum can
be excited in fcc Ni as well as in bcc Nb and Fe
\cite{rate-haas2011,rate-hizhnyakov2014}. Similar results were
obtained for bcc Fe, V, and W~\cite{rate-murzaev2015}.

The point is that the realistic interatomic potentials, including
Lennard-Jones and Morse, have an inflection point meaning that
they are composed of the hard core and the soft tail. This is
typical for interatomic bonds of any complexity, including
many-body potentials. Physically the soft tail is due to the
interaction of the outer electron shells of the atoms, while the
hard core originates from the strong repulsive forces between
nuclei and also from the Pauli exclusion principle for inner
electrons (fermions) that cannot occupy the same quantum state
simultaneously. It is thus important which part of the interatomic
potential (hard or soft) contributes more to the dynamics of the
system. As it was shown in \cite{rate-kiselev1993}, the asymmetry
of the interatomic potentials results in the thermal expansion
effect when larger vibrational amplitudes, at zero pressure, cause
the larger equilibrium interatomic distance and hence, a larger
contribution from the soft tail. If thermal expansion is
suppressed somehow, then the hard core manifests itself. To
demonstrate this let us consider the Morse chain of unit mass
particles whose dynamics is described by the following equations
of motion
\begin{equation} \label{rate-EqMotion}
   \ddot{u}_n=U'(h + u_{n+1}-u_n)-U'(h + u_{n}-u_{n-1})\,,
\end{equation}
where $u_n(t)$ is the displacement of the $n$th particle from the
lattice position, $h$ is the lattice spacing,
\begin{equation} \label{rate-Ur}
   U(r)=D(e^{-2\alpha(r-r_m)}-2e^{- {\alpha} (r-r_m)})\,,
\end{equation}
is the Morse potential, where $r$ is the distance between two atoms,
$D$, $\alpha$, $r_m$ are the potential parameters. The function $U(r)$
has a minimum at $r=r_m$, the depth of the potential (the binding energy)
is equal to $D$ and $\alpha$ defines the stiffness of the bond. We take
$D=1$, $r_m=1$ and $\alpha=5$. For the considered case of the
nearest-neighbor interactions the equilibrium interatomic distance
is $h=r_m=1$.

\begin{figure}[b]
\center
\includegraphics[width=8cm]{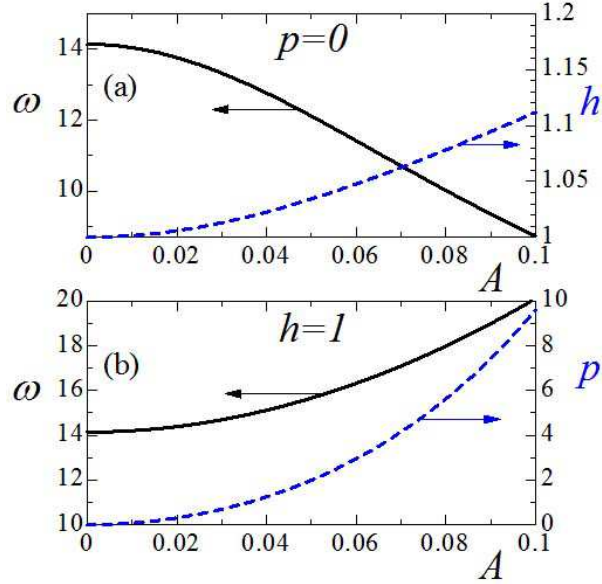} 
\caption{Solid lines show frequency of the staggered mode (left
ordinate) as the function of amplitude for the case of (a) $p=0$
and (b) $h=1$. Dashed lines show (a) $h$ and (b) $p$ (right
ordinate) as the functions of $A$. The results for the 1D Morse
lattice (\ref{rate-EqMotion}), (\ref{rate-Ur}) with the initial
conditions (\ref{rate-StaggMode}).}
 \label{rate-figure01}
\end{figure}

In frame of the model given in Eqs.~(\ref{rate-EqMotion}),
(\ref{rate-Ur}) we study the dynamics of the staggered mode
excited with the use of the following initial conditions
\begin{equation} \label{rate-StaggMode}
   u_n(0)=A\cos(\pi n)=(-1)^n A\,, \quad \dot{u}_n(0)=0\,,
\end{equation}
in the chain of $N$ particles ($N$ is an even number) subjected to
the periodic boundary conditions, $u_n(t)=u_{n+N}(t)$. Our aim is
to find the frequency of the mode as the function of the mode
amplitude $A$ for the two cases. Firstly the chain is allowed to
expand, and for given $A>0$ the interatomic distance $h>1$ is such
that the pressure $p=0$. In the second case the thermal expansion
is suppressed by fixing $h=1$ for any $A$. In this case, of
course, for $A>0$ one has $p>0$. The results for the two cases are
shown in Fig.~\ref{rate-figure01}~(a) and (b), respectively. In
(a) the frequency of the mode decreases with $A$, while in (b) the
opposite takes place.

\rcgindex{\myidxeffect{B}!Breathers above the phonon gap}
\rcgindex{\myidxeffect{B}!Breathers along an atomic row}
\rcgindex{\myidxeffect{A}!Atomic row (breathers along it)}
 In the
numerical experiments by Haas et al. \cite{rate-haas2011} is was
found that the DBs in pure metals are extended along a
close-packed atomic row. The atoms surrounding the atomic row
where DB is excited create the effective periodic on-site
potential that suppresses the thermal expansion of the row and
that is why the DB frequency increases with increasing amplitude.
The on-site potential was not introduced in the 1D model by
Kiselev {\em et al.} \cite{rate-kiselev1993} and, naturally,
thermal expansion did not allow for the existence of DBs with
frequencies above the phonon band.

Notably, the excitation energy of DBs in metals can be
relatively small (fractions of eV) as compared to the formation
energy of a stable Frenkel pair (several eV).
 \rcgindex{\myidxeffect{E}!Energy of breathers in metals}
 \rcgindex{\myidxeffect{M}!Metals (energy of breathers in)}
 \rcgindex{\myidxeffect{B}!Breathers (energy in metals)}
Moreover, it has been shown that DBs in pure metals are highly mobile
and hence they can efficiently transfer energy and momentum over large
distances along close-packed crystallographic directions
\cite{rate-hizhnyakov2014,rate-terentyevdubinko2014,rate-murzaev2015}. 
Recently, a theoretical background has been proposed to ascribe
the interaction of moving DBs (a.k.a 'quodons' - quasi-particles
propagating along close-packed crystallographic directions) with
defects in metals to explain the anomalously accelerated chemical
reactions in metals subjected to irradiation. Russell and Eilbeck
 \cite{rate-russelleilbeck2007}  
  have presented experimental
evidence for the existence of quodons that propagate great
distances in atomic-chain directions in crystals of muscovite, an
insulating solid with a layered crystal structure. Specifically,
when a crystal of muscovite was bombarded with alpha-particles at
a given point at 300 K, atoms were ejected from remote points on
another face of the crystal, lying in atomic chain directions at
more than 10$^7$ unit cells distance from the site of
bombardment. Irradiation may cause continuous generation of DBs
inside materials due to {\em external lattice excitation}, thus
'pumping' a material with DB gas \cite{rate-dubinkodubinko2013,rate-dubinkospringer2014}. 

\begin{figure}[t]
\center
\includegraphics[width=9cm]{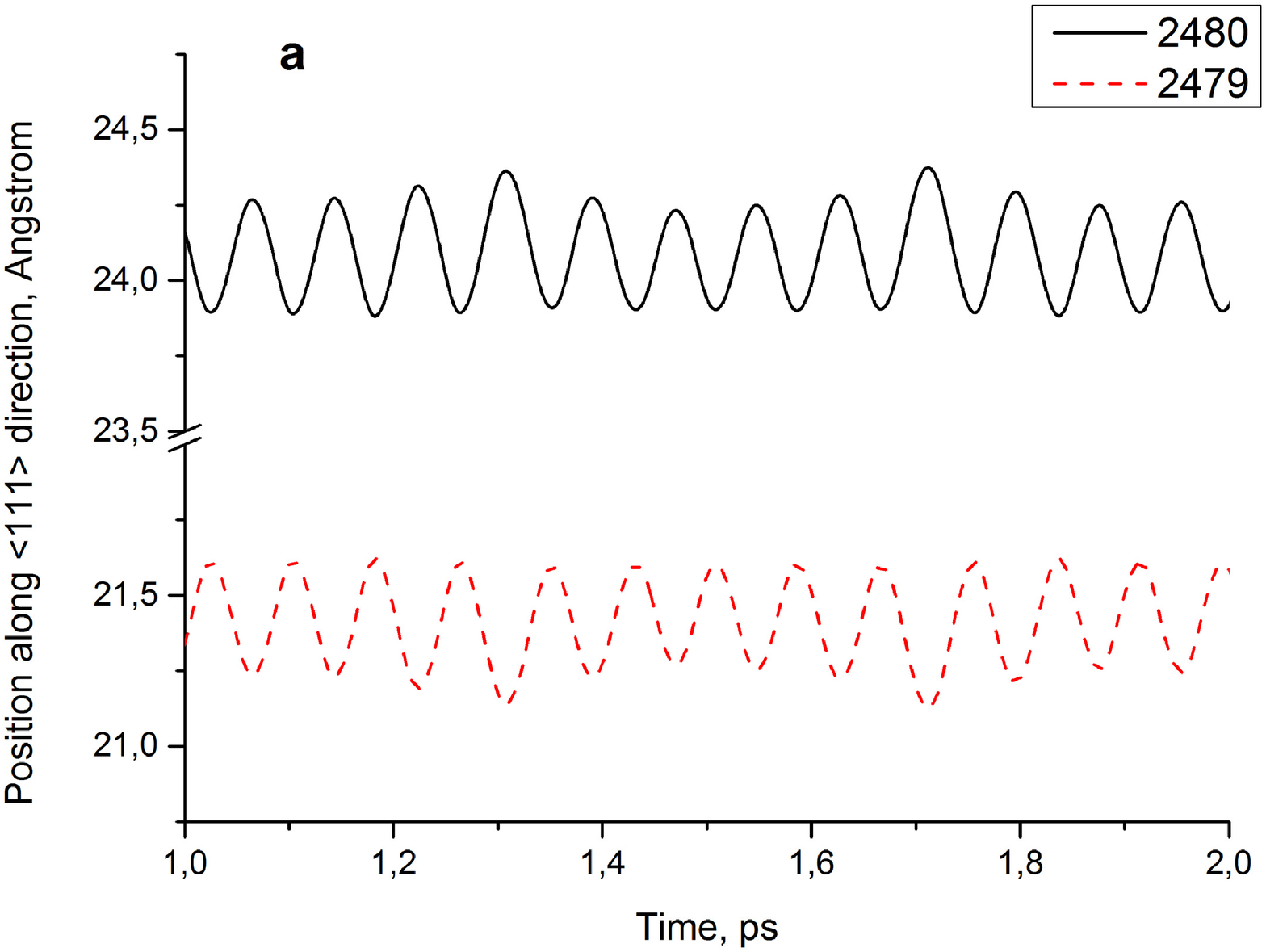}\\
\includegraphics[width=9cm]{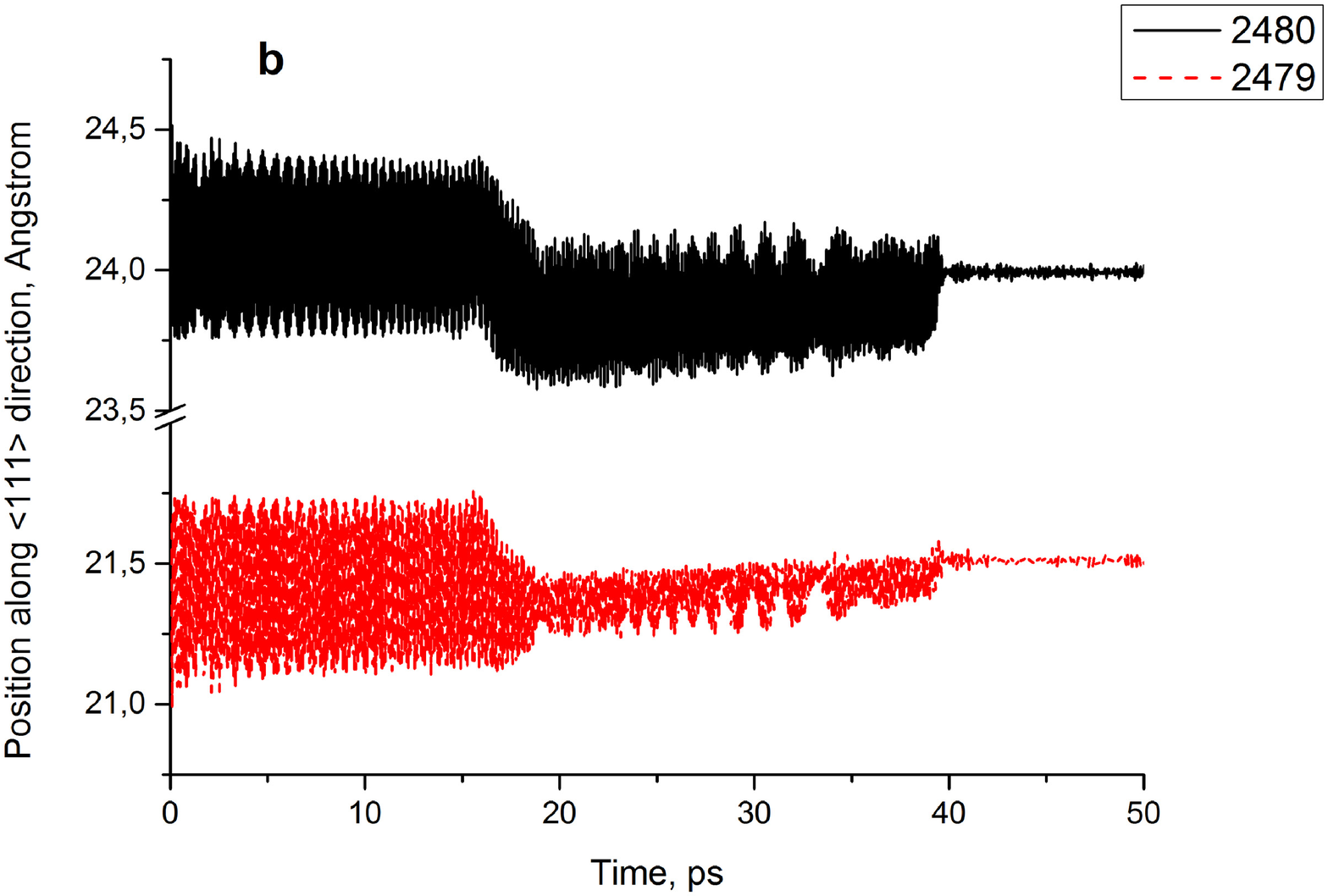}
\caption{Oscillation of $x$ coordinate of two neighbouring atoms,
2480 and 2479, in a [111] row in Fe in a standing DB excited with
$d_0 = 0.325$ \AA  \cite{rate-terentyevdubinko2014}. 
 (a) Initial stage of DB evolution;
(b) total lifespan of DB showing a stepwise quantum nature of its
decay} \label{rate-figure02}
\end{figure}

In order to understand better the structure and properties of
standing and moving DBs, consider the ways of their external
excitation in Fe by MD simulations  \cite{rate-terentyevdubinko2014}. 
A standing DB can be excited by applying the initial displacements
to the two adjacent atoms along the close-packed [111] direction
with the opposite signs to initiate their {\em anti-phase}
oscillations, as shown in Fig. \ref{rate-figure02}(a). The initial
displacements $\pm d_0$ determine the DB amplitude, frequency and,
ultimately, its lifetime. DBs can be excited in a frequency band
(1.0-1.4)$\times 10^{13}$~Hz just above the Debye frequency of bcc
Fe, and DB frequency grows with increasing amplitude as expected
for the hard type anharmonicity due to the major contribution from
the hard core of the interatomic potential. Initial displacements
larger than $|d_0|=0.45$~\AA~ generate a chain of {\em focusons},
while displacements smaller than $|d_0|=0.27$~\AA~ do not provide
enough potential energy for the system to initiate a stable DB and
the atomic oscillations decay quickly by losing its energy to {\em
phonons}. The most stable DBs can survive up to 400 oscillations,
as shown in Fig.~\ref{rate-figure02}(b), and ultimately decay in a
stepwise quantum nature by generating bursts of phonons, as has
been predicted by Hizhnyakov as early
as in 1996 \cite{rate-hizhnyakov1996}. 
\rcgindex{\myidxeffect{D}!Decay of breathers}
\rcgindex{\myidxeffect{B}!Breathers (decay)}
\rcgindex{\myidxeffect{B}!Breathers (lifetime)}
 \rcgindex{\myidxeffect{L}!Lifetime of breathers}
\begin{figure}[t]
\center
\includegraphics[width=8cm]{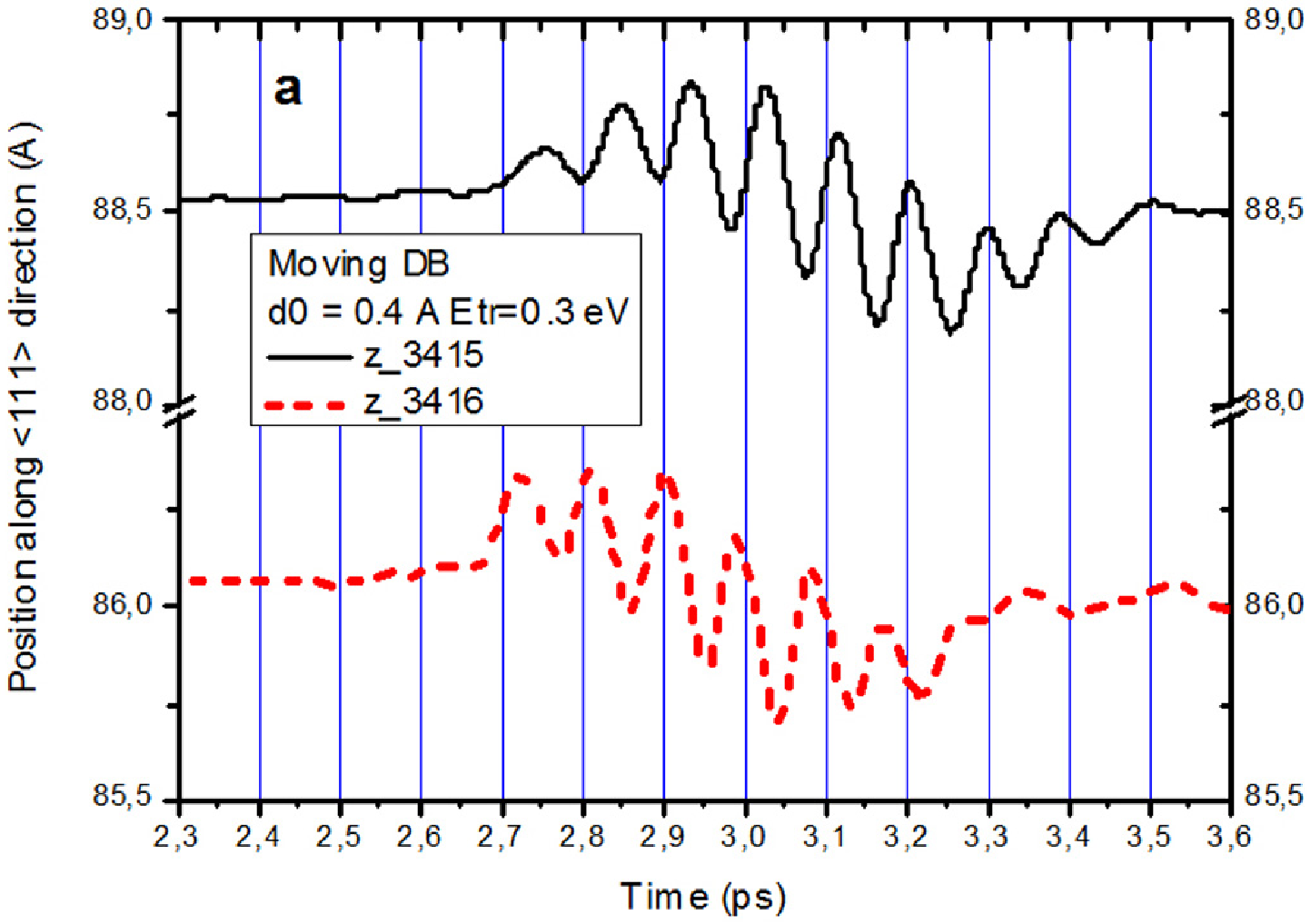} %
\includegraphics[width=8cm]{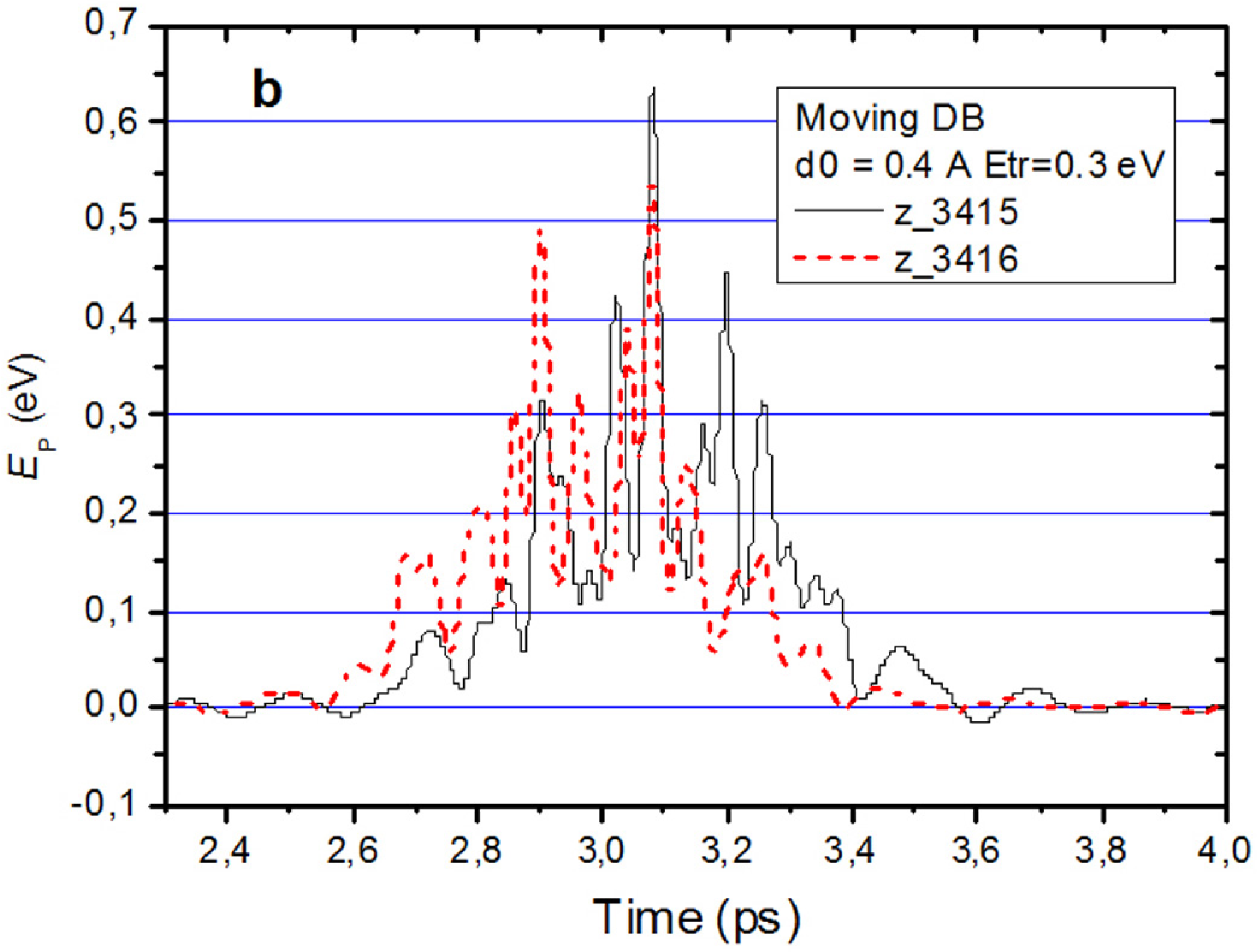}
 \caption{(a) Oscillation
of $x$ coordinate of two neighbouring atoms, 3415 and 3416 in a
[111] row in Fe during the passage of a moving DB ($d_0 = 0.4$
\AA, $E_{tr}= 0.3$ eV); (b) deviation of the potential energy of
the atoms from the ground state during the passage of DB}
\label{rate-figure03}
\end{figure}

A moving DB can be excited by introducing certain asymmetry into
the initial conditions. Particularly, the translational kinetic
energy $E_{tr}$ can be given to the two central atoms of DB in the
same direction along [111] atomic row. DB velocity ranges from 0.1
to 0.5 of the velocity of sound, while travel distances range from
several dozens to several hundreds of the atomic spaces, depending
on $d_0$ and $E_{tr}$
\cite{rate-terentyevdubinko2014,rate-murzaev2015}.
Figure~\ref{rate-figure03}(a) shows a DB passing the two
neighboring atoms with indices 3415 and 3416. In the moving DB the
two central atoms pulsate not exactly in anti-phase but with a
phase shift.
  \rcgindex{\myidxeffect{B}!Breather mean free path}
    \rcgindex{\myidxeffect{M}!Mean free path of breathers}
In about 1~ps ($\sim$10 oscillations) the oscillations of these
two atoms cease but they are resumed at the subsequent atoms along
[111] atomic row. In this way, the DB moves at a speed of 2.14
km/s, i.e. about the half speed of sound in bcc Fe. The
translational kinetic energy of the DB is about 0.54~eV, which is
shared mainly among two core atoms, giving 0.27~eV per atom, which
is close to the initial kinetic energy of $E_{tr}=0.3$~eV given to
the atoms to initiate the DB translational motion. The deviation
of the potential energy of the atoms from the ground state during
the passage of the DB is presented in Fig.~\ref{rate-figure03}(b).
The maximal deviation of energy is of the order of 1~eV. Thus, a
moving DB can be viewed as an atom-size localised excitation with
local temperature above 1000~K propagating along the crystal at a
subsonic speed.

\begin{figure}[t]
 \center
\includegraphics[width=8cm]{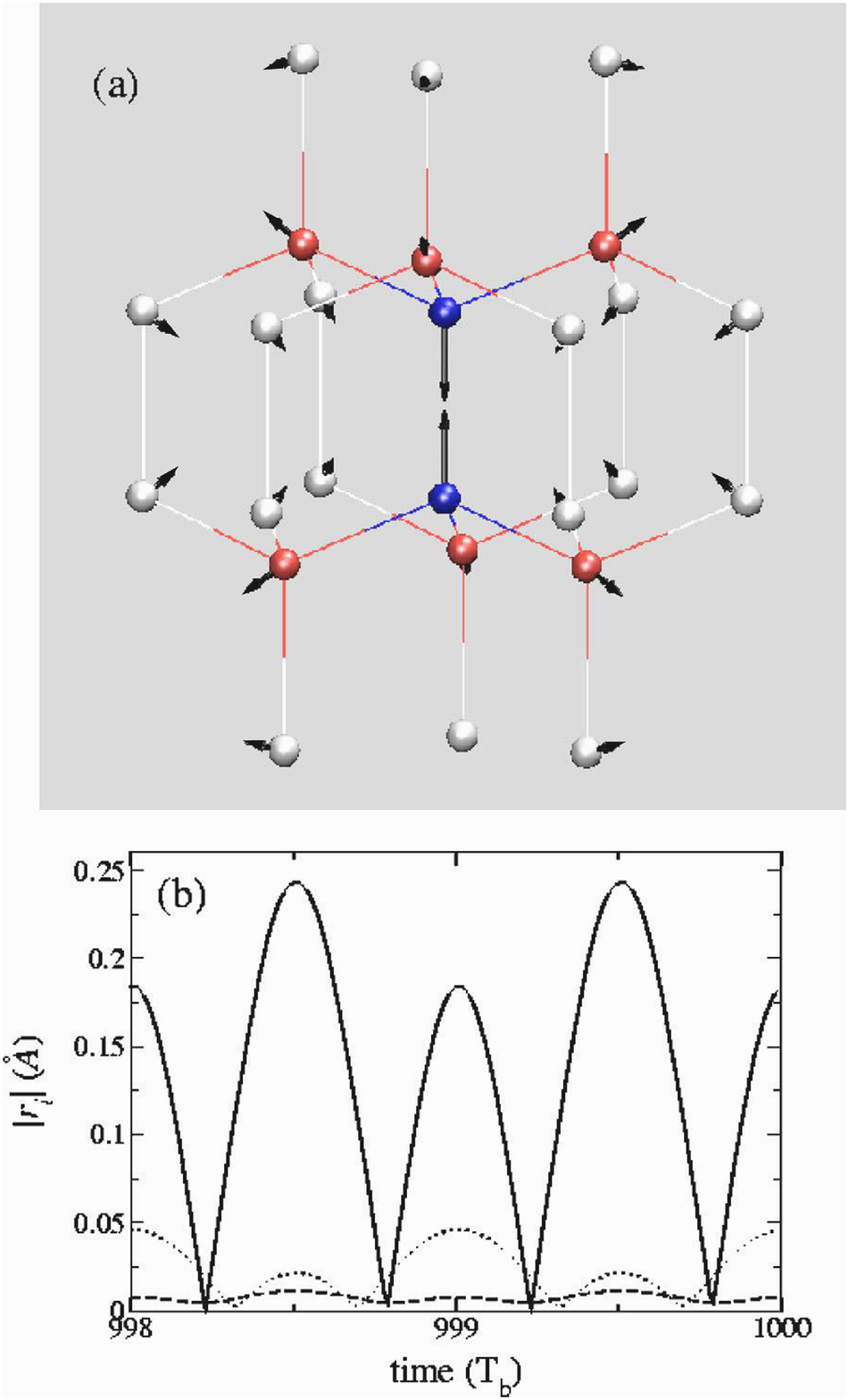}
\caption{(a) DB generation in silicon modeled by Tersoff
potentials. The DB frequency is $1.733\times 10^{13}$ Hz, while
vectors (magnified for visualization purposes) denote atomic
displacements from equilibrium; only first (gray, red online) and
second (white) neighbors to the central (black, blue online) two
breather atoms are included. The displacement of the two central
breather atoms is 0.18 \AA. (b) Time evolution of the silicon DB
after 998 breather periods. The absolute value of the
displacements from equilibrium along the direction of motion of
each atom is plotted. The coordinated oscillations of central
(solid), first (dotted), and second (dashed) neighbor atoms are
indicated. Reproduced with permission from Voulgarakis, N.,
Hadjisavvas, G., Kelires, P., Tsironis, G.: Computational
  investigation of intrinsic localization in crystalline {Si}.
Phys. Rev. B \textbf{69}, 113,201 (2004). Copyright (2004)
American Physical Society}
 \label{rate-figure04} 
\end{figure}

\subsection{Semiconductors}  \label{subsec:dubinko:semiconductors} 
       \rcgindex{\myidxeffect{B}!Breathers in semiconductors}
         \rcgindex{\myidxeffect{S}!Semiconductor (breathers in)}
Similar to metals, semiconductors possess no gap in phonon
spectrum and thus DBs may exist only if their frequency is
positioned above the phonon spectrum
\cite{rate-voulgarakis2004,rate-haas2011}. Such high-frequency DBs
may be realized in semiconductors due to the screening of the
short-range covalent interaction by the conducting electrons.
Voulgarakis et al. \cite{rate-voulgarakis2004} investigated
numerically existence and dynamical properties of DBs in
crystalline silicon through the use of the Tersoff interatomic
potential. They found a band of DBs with lifetime of at least
60~ps in the spectral region $(1.643-1.733)\times 10^{13}$~Hz,
located just above the upper edge of the phonon band calculated at
$1.607\times 10^{13}$~Hz. The localized modes extend to more than
second neighbors and involve pair central-atom compressions in the
range from 6.1\% to 8.6\% of the covalent bond length per atom.
Finite temperature simulations showed that they remain robust to
room temperatures or higher with a typical lifetime equal to 6~ps.
Figure~\ref{rate-figure04} shows DB generated in silicon modeled
by the Tersoff potential \cite{rate-voulgarakis2004}. It can be
seen that the DB is very persistent and localized: its vibrational
energy is mainly concentrated in the bond between two neighboring
atoms oscillating in anti-phase mode.

\begin{figure}[t]
\center
\includegraphics[width=8cm]{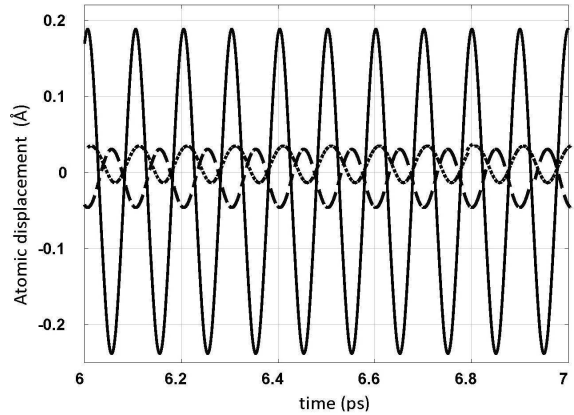}
\caption{DB generated in germanium modeled by the Tersoff
potential. Displacement of one of the two central atoms is shown
with a solid line and of the first neighbor by dashed (along [111]
axis) and dotted (perpendicular to [111] axis) lines. See
Ref.~\cite{rate-hizhnyakovquodons2015}}
 \label{rate-figure05} 
\end{figure}

Similar to silicon, germanium has a diamond crystal structure and
readily produces DBs \cite{rate-hizhnyakovquodons2015}, as
demonstrated in Fig.~\ref{rate-figure05}. As in Si, the DB's
energy in Ge is concentrated in the central bond between two atoms
oscillating in anti-phase mode. This means that potential barriers
for chemical reactions in the vicinity of an DB may be subjected
to persistent periodic oscillations, which has been shown to
result in a strong amplification of the reaction rates
 \cite{rate-dubinkoarchilla2011}.
         \rcgindex{\myidxeffect{R}!Reaction rate amplification by DB}
In the next section we consider the ways of DB excitation in thermal
equilibrium and under external driving.

\section{DB excitation under thermal equilibrium and external driving}
  \label{sec:dubinko:DBexcitation}   

In this section, for the convenience of the reader, we repeat the
main points of the chemical reaction rate theory that takes into
account the effect of DBs, following the earlier works
\cite{rate-archillacuevas2006,rate-dubinkoarchilla2011,rate-dubinko2014}.
       \rcgindex{\myidxeffect{R}!Reaction rate theory and breathers}
       \rcgindex{\myidxeffect{B}!Breathers and reaction rate theory}
 \begin{figure}[t]
\center
\includegraphics[width=\textwidth]{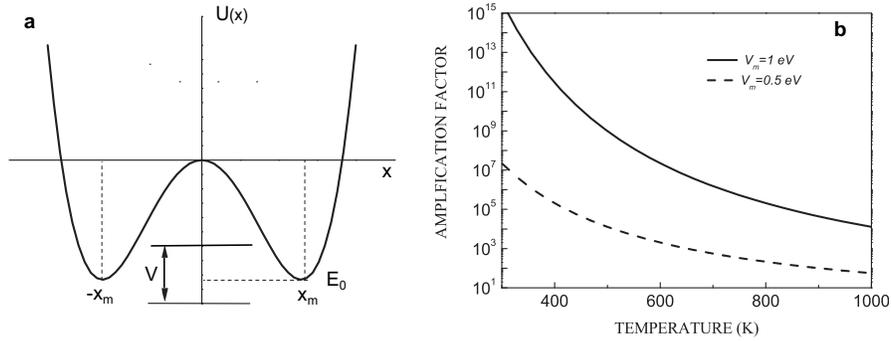}
\caption{(a) Sketch of the double-well potential landscape with
minima located at $\pm x_m$. These are stable states before and
after reaction, separated by a potential "barrier" with the height
changing periodically or stochastically within the $V$ band. (b)
Amplification factor, $I_0(V/k_B T)$, for the average escape rate
of a thermalized Brownian particle from a periodically modulated
potential barrier at different temperatures and modulation
amplitudes $V$. Reproduced with permission from V.I. Dubinko, P.A. Selyshchev and J.F.R. Archilla:
 Reaction-rate theory with account of the crystal anharmonicity
 \newblock Phys. Rev. E \textbf{83} 041124 (2011).
 Copyright (2011) American Physical Society.}
\label{rate-figure06}
\end{figure}

The rate equation for the concentration of DBs with energy $E$,
$C_{DB}(E,t)$ can be written as follows
\cite{rate-dubinkoarchilla2011}
\begin{eqnarray}
    \label{rate-eq:dubinko:1}
    \frac{\partial C_{DB}(E,t)}{\partial t}=K_{DB}(E)-\frac{
    C_{DB}(E,t)}{\tau_{DB}(E)},
\end{eqnarray}
where $K_{DB}(E)$ is the rate of creation of DBs with energy
$E>E_{\min}$ and $\tau_{DB}(E)$ is the DB lifetime. It has an
obvious steady-state solution ($\partial C_{DB}(E,t)/\partial
t=0$):
\begin{eqnarray}
    \label{eq:dubinko:2}
    C_{DB}(E)=K_{DB}(E)\tau_{DB}(E).
\end{eqnarray}
In the following sections we will consider the breather formation
by thermal activation and then extend the model to non-equilibrium
systems with external driving.

\subsection{Thermal activation}  \label{subsec:dubinko:thermal} 

The exponential dependence of the concentration of high-energy
light atoms on temperature in the MD simulations
\cite{rate-khadeeva2011} gives evidence in favor of their thermal
activation at a rate given by a typical Arrhenius law
\cite{rate-piazza2003}
\rcgindex{\myidxeffect{T}!Thermal activation of breathers}
\rcgindex{\myidxeffect{B}!Breathers (thermal activation of)}
\begin{eqnarray}
    \label{eq:dubinko:3}
    K_{DB}(E,T)=\omega_{DB}\exp\left(-\frac{E}{k_BT} \right),
\end{eqnarray}
where $\omega_{DB}$ is the attempt frequency that should be close
to the DB frequency. The breather lifetime has been proposed in
\cite{rate-piazza2003} to be determined by a phenomenological law
based on fairly general principles: (i) DBs in two and three
dimensions have a minimum energy $E_{\min}$, (ii) The lifetime of
a breather grows with its energy as
$\tau_{DB}=\tau^0_{DB}(E/E_{\min}-1)^z$, with $z$ and
$\tau^0_{DB}$ being constants, whence it follows that under
thermal equilibrium, the DB energy distribution function
$C_{DB}(E,T)$ and the mean number of breathers per site
$n_{DB}(T)$ are given by
\begin{eqnarray}
    \label{eq:dubinko:4}
    C_{DB}(E,T)=\omega_{DB}\tau_{DB}\exp\left(-\frac{E}{k_BT} \right),
\end{eqnarray}
\begin{eqnarray}
    \label{eq:dubinko:5}
    n_{DB}(T)=\int\limits_{E_{\min}}^{E_{\max}}C_{DB}(E,T)dE =
    \omega_{DB}\tau^0_{DB}\frac{\exp\left(-E_{\min}/{k_BT}
    \right)}{(E_{\min}/{k_BT})^{z+1}}
    \int\limits_0^{y_{\max}}y^z\exp(-y)dy\,,
\end{eqnarray}
 with $y_{\max}=(E_{\max}-E_{\min})/{k_B T}$. Noting that $\Gamma(z+1,x)=\int_0^xy^z\exp(-y)dy$ is the second
incomplete gamma function, Eq.~(\ref{eq:dubinko:5}) can be written
as \cite{rate-dubinkoarchilla2011}:
\begin{eqnarray}
    \label{eq:dubinko:6}
    n_{DB}(T)=\omega_{DB}\tau^0_{DB}\frac{\exp\left(-E_{\min}/k_BT
    \right)}{(E_{\min}/k_BT)^{z+1}}
    \Gamma\left( z+1, \frac{E_{\max}-E_{\min}}{k_BT}\right).
\end{eqnarray}
It can be seen that the mean DB energy is higher than the averaged
energy density (or temperature):
\begin{eqnarray}
    \label{eq:dubinko:7}
    \langle E_{DB}\rangle=\frac{\int\limits_{E_{\min}}^{E_{\max}}C_{DB}(E,T)EdE}{\int\limits_{E_{\min}}^{E_{\max}}C_{DB}(E,T)dE}
    \overrightarrow{_{E_{\max}\gg E_{\min}}}\left(\frac{E_{\min}}{k_BT}+z+1 \right)\times k_BT.
\end{eqnarray}

Assuming, according to \cite{rate-khadeeva2011}
  that $E_{\min}/k_BT \approx 3$ and
$\langle E_{DB}\rangle\approx 5k_BT$, one obtains an estimate for
$z\approx 1$, which corresponds to linear increase of the DB
lifetime with energy.

\subsection{External driving}  \label{subsec:dubinko:driving} 

\rcgindex{\myidxeffect{E}!Externally driven breathers}
\rcgindex{\myidxeffect{B}!Breathers with external driving}
 Fluctuation activated nature
of DB creation can be described in the framework of classical
Kramers model, which is archetypal for investigations in
reaction-rate theory \cite{rate-hanggi1990}. The model considers a
Brownian particle moving in a symmetric double-well potential
$U(x)$ (Fig. \ref{rate-figure06}(a)). The particle is subject to
fluctuational forces that are, for example, induced by coupling to
a heat bath The fluctuational forces cause transitions between the
neighboring potential wells with a rate given by the famous
Kramers rate:
\begin{eqnarray}
    \label{eq:dubinko:8}
    \dot{R}_K(E_0,T)=\omega_0\exp(-E_0/ k_BT),
\end{eqnarray}
where $\omega_0$ is the attempt frequency and $E_0$ is the height
of the potential barrier separating the two stable states, which,
in the case of fluctuational DB creation, corresponds to the
minimum energy that should be transferred to particular atoms in
order to initiate a stable DB. Thus, the DB creation rate (3) is
given by the Kramers rate: $K_{DB}(E,T)=\dot{R}_K(E,T)$.

In the presence of {\em periodic modulation} (driving) of the well
depth (or the reaction barrier height) such as
$U(x,t)=U(x)-V(x/x_m)\cos(\Omega t)$, the reaction $\dot{R}_K$
rate averaged over times exceeding the modulation period has been
shown to increase according to the following equation
\cite{rate-dubinkoarchilla2011}:
\begin{eqnarray}
    \label{eq:dubinko:9}
    \langle \dot{R}\rangle_m=\dot{R}_KI_0\left(\frac{V}{k_BT}\right),
\end{eqnarray}
where the amplification factor $I_0(x)$ is the zero order modified
Bessel function of the first kind. Note that the amplification
factor is determined by the ratio of the modulation amplitude $V$
to temperature, and it does not depend on the modulation frequency
or the mean barrier height. Thus, although the periodic forcing
may be too weak to induce {\em athermal} reaction (if $V<E_0$), it
can amplify the average reaction rate drastically if the ratio
$V/k_BT$ is high enough, as it is demonstrated in Fig.
\ref{rate-figure06}(b).
    \rcgindex{\myidxeffect{A}!Activation barriers and DBs}
     \rcgindex{\myidxeffect{B}!Breathers and activation barriers}
Another mechanism of enhancing the DB creation rate is based on
small {\em stochastic modulations} of the DB activation barriers
caused by external driving. Stochastic driving has been shown to
enhance the reaction rates via effective reduction of the
underlying reaction barriers
 \cite{rate-dubinkodubinko2013,rate-dubinkospringer2014} 
 as:
\begin{eqnarray}
    \label{eq:dubinko:10}
    \langle \dot{R}\rangle=\omega_0\exp(-E_a^{DB}/k_BT), \quad
    E_a^{DB}=E_0-\frac{\langle V\rangle^2_{SD}}{2k_BT}, \quad
    {\rm if}\,\,\,\,\langle V\rangle_{SD}\ll k_BT,
\end{eqnarray}
where $\langle V\rangle_{SD}$ is the standard deviation of the
potential energy of atoms surrounding the activation site.

In the present view, the DB creation is seen as a {\em chemical
reaction} activated by thermally or externally induced
fluctuations. In the following section we consider the reaction of
annealing of defects in crystals, such as the deep traps for
electrons/holes, within the similar framework. I simplified model
can be seen in Ref.~\cite{rate-coelhoarchillaquodons2015}

\section{Amplification of Sb-vacancy annealing rate in germanium by DBs}
  \label{sec:dubinko:DBamplification} 

\begin{figure}[t]
\center
\includegraphics[width=\textwidth,clip]{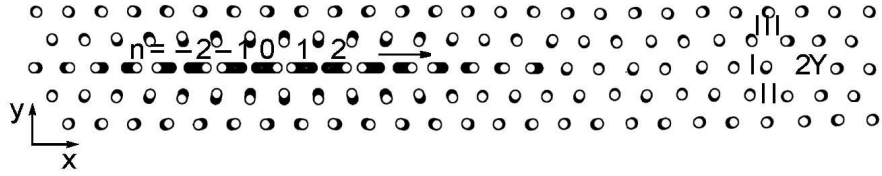}
\caption{Illustration of moving DB (quodon) before "collision"
with a vacancy in 2D crystal (4 times zoom of atomic
displacements) \cite{rate-kistanov2014}. $2Y$ is the distance
between the atoms II and III.  Reproduced with permission from
Kistanov, A., Dmitriev, S., Semenov, A.S., Dubinko, V., Terentyev,
D.: Interaction of propagating discrete breathers with a vacancy
in a two-dimensional crystal. Tech. Phys. Lett. \textbf{40},
657–661 (2014). Copyright (2014) Springer.}
\label{rate-figure07}
\end{figure}
      \rcgindex{\myidxeffect{A}!Annealing of defects by DBs}
         \rcgindex{\myidxeffect{B}!Breathers that anneal defects}
\begin{figure}[t]
\center
\includegraphics[width=10cm]{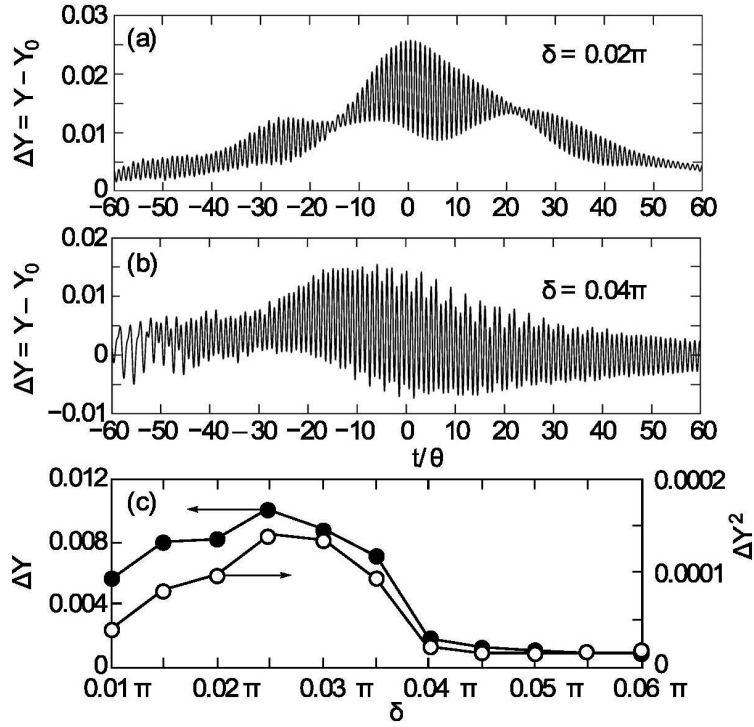}
\caption{(a)  Dependence of $\Delta Y=Y-Y_0$ on time for "slow"
DBs (b) "fast" DBs ; (c) Mean difference $\langle \Delta Y\rangle$
and standard deviation over 80 oscillation periods vs. phase
difference $\langle \Delta Y \rangle^2$, which is proportional to
the DB velocity. It can be seen that "slow" DBs disturb the
vacancy more strongly than the "fast" ones
\cite{rate-kistanov2014}. Reproduced with permission from
Kistanov, A., Dmitriev, S., Semenov, A.S., Dubinko, V., Terentyev,
D.: Interaction of propagating discrete breathers with a vacancy
in a two-dimensional crystal.
\newblock Tech. Phys. Lett. \textbf{40}, 657–661 (2014).
Copyright (2014) Springer.}
\label{rate-figure08}
\end{figure}
           \rcgindex{\myidxeffect{E}!E-center defect}
             \rcgindex{\myidxeffect{D}!Defect E-center}
           \rcgindex{\myidxeffect{S}!Sb-vacancy complex}
   \rcgindex{\myidxeffect{V}!Vacancy-Sb complex}
    \rcgindex{\myidxeffect{E}!Electron trap}
     \rcgindex{\myidxeffect{T}!Trap of electrons}
Sb-vacancy defect in Ge is a typical deep trap, which has been
shown to arise under displacement damage (producing vacancies) and
anneal either thermally (above 400~K) or under ICP treatment at
ambient temperatures of about 300~K
\cite{rate-archillacoelho2015}. This plasma-induced acceleration
of annealing at depth extending up to several microns must be
driven by some mechanism capable of transferring the excitation
energy of surface atoms (interacting with plasma) deep into the
crystal. Quodons are thought to be good candidates for providing
such a mechanism, and bellow we present a model of quodon-enhanced
defect annealing based on quasi-periodic modulation of the
annealing activation barrier caused by the interaction of defects
with a 'quodon gas'. This mechanism is illustrated in
Fig.~\ref{rate-figure07}, which shows a moving DB (quodon) before
'collision' with a vacancy in 2D close-packed crystal with
pairwise Morse interatomic potentials \cite{rate-kistanov2014}.
The DB velocity can be varied by changing the phase difference,
$\delta$. The distance between the atoms II and III is $2Y$ and
$\Delta Y=Y-Y_0$ is the difference between the excited and ground
state due to the interaction with a quodon, which is shown in Fig.
\ref{rate-figure08} as a function of time for 'slow' and 'fast'
DBs. The mean difference $\langle \Delta Y\rangle$ and standard
deviation $\langle \Delta Y\rangle^2$ over the excitation time of
$\sim 80$ oscillation periods have been calculated. It can be seen
that "slow" DBs disturb the vacancy more strongly than the "fast"
ones, and besides, they practically do not lose their energy in
the course of 'collision'. So these DBs behave similar to
molecules of some gas, which can be 'pumped' from the surface into
material up to some depth equal to the propagation range of
quodons before the decay. Then, the average rate of quodon
generation (per atom), will be proportional to the ratio of their
flux $\Phi_q$ though the surface (where they are created by
energetic plasma atoms) to the propagation range of quodons,
$l_q$:
            \rcgindex{\myidxeffect{P}!Plasma (generation of DB by)}
         \rcgindex{\myidxeffect{G}!Generation of DB by plasma}
  \rcgindex{\myidxeffect{B}!Breathers generated by plasma}
\begin{eqnarray}
    \label{eq:dubinko:11}
    K_q=\frac{\Phi_q}{l_q}\omega_{\rm Ge}, \quad
    \Phi_q=\Phi_{\rm Ar}\frac{4E_{\rm Ar}M_{\rm Ar}M_{\rm Ge}}{E_q(M_{\rm Ar}+M_{\rm Ge})^2},
\end{eqnarray}
where $\omega_{\rm Ge}$ is the Ge atomic volume, $M_{\rm Ar}$,
$M_{\rm Ge}$ are the Ar and Ge atomic masses, $\Phi_{\rm Ar}$ is
the flux of Ar ions or atoms with a mean energy $E_{\rm Ar}$, a
part of which $4M_{\rm Ar}M_{\rm Ge}/(M_{\rm Ar}+M_{\rm Ge})^2$,
is transferred to germanium atoms and could be spent on the
generation of quodons with a mean energy $E_q$. Then the
steady-state concentration of quodon gas (see Fig.
\ref{rate-figure09}) will be given simply by the product of their
generation rate and the life-time, $\tau_q$:
\begin{eqnarray}
    \label{eq:dubinko:12}
    C_q=K_q\tau_q, \quad \tau_q=\frac{l_q}{v_q}, \quad {\rm then}
    \quad C_q=\frac{\Phi_q\omega_{\rm Ge}}{v_q},
\end{eqnarray}
where $v_q$ is the quodon propagation speed, which actually
determines their concentration within the layer of a thickness
$l_q$ (Fig. \ref{rate-figure09}).

\begin{figure}[t]
\center
\includegraphics[width=8cm]{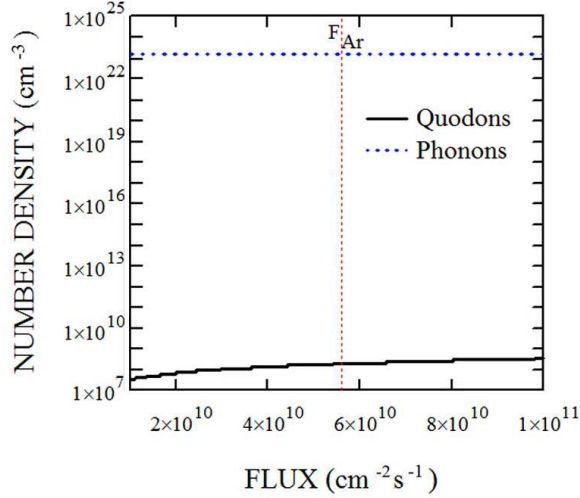}
\caption{The density of quodon gas, $C_q/\omega_{Ge}$, vs. Ar flux
at the irradiation temperature of 300 K  within the layer of
thickness $L_q = 5.3$ microns, at the quodon velocity of $v_q=
300$ m/s. Density of the phonons at 300 K is shown for comparison
with a dotted line. The vertical dotted line corresponds to Ar
flux in the experiment \cite{rate-archillacoelho2015}.}
\label{rate-figure09}
\end{figure}
    \rcgindex{\myidxeffect{Q}!Quodon gas}
        \rcgindex{\myidxeffect{G}!Gas of quodons}
\begin{figure}[t]
\center
\includegraphics[width=\textwidth]{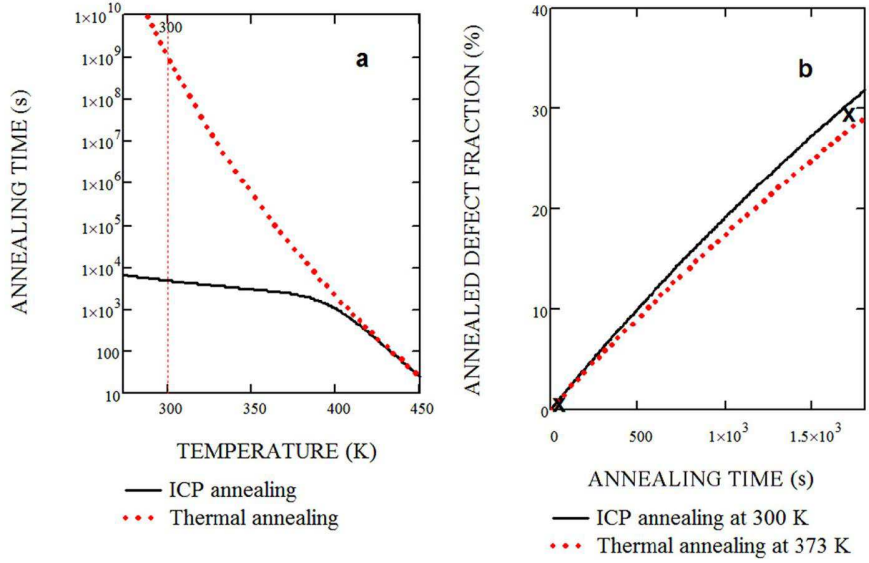}
\caption{(a) Characteristic annealing time,
Eq.~(\ref{eq:dubinko:16}) under thermal treatment and ICP. (b)
Annealed defect fraction with time during thermal annealing at 373
K  in comparison with ICP-induced annealing at 300 K according to
the Eq.~(\ref{eq:dubinko:15})  and experimental data X.
Irradiation and material parameters: $F_{Ar}=5.6\times 10^{10}$
cm$^{-2}$s$^{-1}$; $\tau_{ex}=10^{-11}$s; $\omega_0=5.313\times
10^{13}$s$^{-1}$; $E_a=1.35$ eV; $V_{ex}=1.28$ eV. }
\label{rate-figure10}
\end{figure}

\begin{figure}[t]
\center
\includegraphics[width=\textwidth]{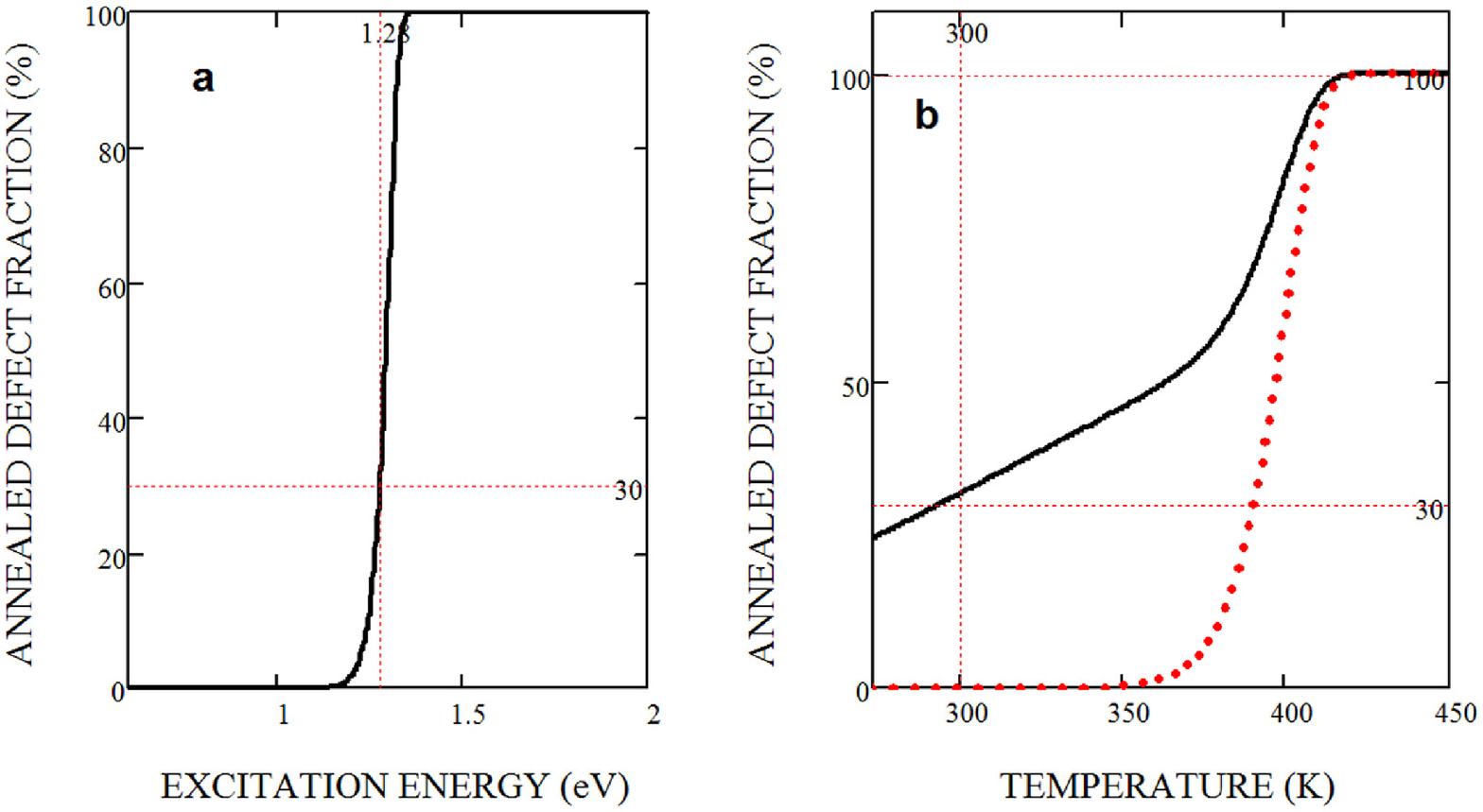}
\caption{(a) Annealed defect fraction at after 30 min of ICP vs.
excitation energy, $V_{ex}$, at 300 K ; (b) after 30 min of ICP or
heating vs. temperature at  $V_{ex}$=1.28 eV.}
\label{rate-figure11}
\end{figure}

Consider the periodic modulation of the defect annealing
activation energy in more details. It is driven by quodons that
scatter on the defects and excite the surrounding atoms (Fig.
\ref{rate-figure08}). The amplitude of the quasi-periodic energy
deviation $V_{ex}$ can be in the eV range with the excitation
time, $\tau_{ex}$, of about 100 oscillation periods. In the
modified Kramers model (\ref{eq:dubinko:9}), this energy deviation
corresponds to the modulation of the annealing activation barrier.
Then, a {\em macroscopic} annealing rate (per defect per second)
may be written as follows:
\begin{eqnarray}
    \label{eq:dubinko:13}
    \langle \dot{R}\rangle_{macro}=\omega_0\exp\left( -\frac{E_a}{k_BT}\right)
    \left(1+\Big\langle I_0\frac{V_{ex}}{k_BT}\Big\rangle\omega_{ex}\tau_{ex}\right),
\end{eqnarray}
where $E_a$ is the annealing activation energy, $\omega_{ex}$ is
the mean number of excitations per defect per second caused by the
flux of quodons, which is proportional to the quodon flux and the
cross-section of quodon-defect interaction and is given by
\begin{eqnarray}
    \label{eq:dubinko:14}
    \omega_{ex}=\Phi_q\pi b^2,
\end{eqnarray}
where $b$ is the atomic spacing, the quodon formation energy
$V_q\approx V_{ex}$. For material parameters presented in Fig.
\ref{rate-figure10}, one has $\omega_{ex} \approx
10^{-4}$s$^{-1}$.

Sb-vacancy annealing kinetics is described by the following
equation for the defect concentration:
\begin{eqnarray}
    \label{eq:dubinko:15}
    \frac{dc_d}{dt}=-\frac{c_d}{\tau_a}, \quad c_d(t)=c_d(0)\exp\left(
    -\frac{t}{\tau_a}\right),
\end{eqnarray}
where $\tau_a$ is the characteristic annealing time, which
inversely proportional to the annealing reaction rate given by
Eq.~(\ref{eq:dubinko:13})
\begin{eqnarray}
    \label{eq:dubinko:16}
    \tau_a=\frac{\exp\left( \frac{E_a}{k_BT}\right)}{\omega_0\left(1+I_0 \left(\frac{E_{ex}}{k_BT} \right)\omega_{ex}\tau_{ex} \right)}.
\end{eqnarray}

In the absence of driving ($\Phi_{\rm Ar}=0 => \omega_{ex}=0$),
Eq.~(\ref{eq:dubinko:16}) describes the {\em thermal annealing},
while at $\Phi_{\rm Ar}>0$, the annealing proceeds at room
temperatures at a rate which is 5 orders of magnitude higher than
that at $\Phi_{\rm Ar}=0$, and it is comparable to the thermal
annealing at the boiling point (373~K), as demonstrated in Fig.
\ref{rate-figure10}. In agreement with experimental data
\cite{rate-archillacoelho2015}, the defect concentration decreases
by 30\% after ICP treatment for 30~min at room temperature.

The ICP-annealing rate is very sensitive to the excitation energy
(Fig. \ref{rate-figure11}(a)), and it increases monotonously with
temperature (Fig. \ref{rate-figure11}(b)), provided that the
quodon production rate and propagation range are temperature
independent.

\section{Summary}
 \label{sec:dubinko:summary} 

A new mechanism of the long-range annealing of defects in Ge under
low energy ICP treatment is proposed, which is based on the
catalyzing effect of DBs on annealing reactions.  The moving DB
(quodon) creation is triggered by Ar flux which provides the input
energy transformed into the lattice vibrations.

Simple analytical expressions for the annealing rate under ICP
treatment are derived as functions of temperature, ion current and
material parameters, which show a good agreement with experimental
data.

\section*{Acknowledgments} S.V.D. thanks the Tomsk State University Academic D.I. Mendeleev Fund Program.
\\ \mbox{}\\
\bibliographystyle{spmpsci}
\bibliography{dbrate}
\end{document}